\documentclass[useAMS]{mn2e}

\usepackage[dvips]{graphicx}
\usepackage{amssymb}
\usepackage{url}

\title[Constraints on GRB external shock prior emission]{Observational constraints on the external shock prior emission hypothesis of GRBs}
\author[T. Birnbaum, B. Zhang, B.-B. Zhang and E.-W. Liang]{Tesla Birnbaum\thanks{E-mail: tesla.birnbaum@gmail.com}$^{1}$, Bing Zhang$^{1}$, Bin-Bin Zhang$^{2}$ and En-Wei Liang$^{3}$\\
$^{1}$Department of Physics and Astronomy, University of Nevada Las Vegas, Las Vegas, NV 89154, USA\\
$^{2}$Department of Astronomy and Astrophysics, Pennsylvania State University, University Park, PA 16802, USA\\
$^{3}$Department of Physics, Guangxi University, Nanning 530004, China}
\begin{document}

\date{Accepted 2012 January 22. Received 2012 January 22; in original form 2011 May 24}

\pagerange{\pageref{firstpage}--\pageref{lastpage}} \pubyear{2011}

\maketitle

\label{firstpage}

\begin{abstract}
An intriguing hypothesis, that there exists a decaying X-ray emission component before the GRB trigger, has been suggested in order to explain the shallow decay phase of the X-ray afterglow detected in many {\em Swift} GRBs. If this ``prior emission'' is from an external shock, one would expect a corresponding optical emission component during the GRB prompt emission phase. In this paper we apply the available prompt optical emission data (both detections and upper limits) to constrain such a scenario. We fit the shallow and normal decay segments of the XRT light curves in our sample with a $T_\Delta$-shifted single power law, and extrapolate the X-ray flux back to the time of the early optical observations. We then use the synchrotron spectrum predicted by the standard external shock model to extrapolate from the X-ray flux to the optical band, and obtain the possible range of the predicted optical flux. Finally, we compare the predictions with the observations.  In the cases where later optical data are available, we also compare the shapes of the optical light curves to the predicted optical light curves from the external shock prior emission model. We find that for a good fraction of GRBs (4 out of 8; up to December 2006), the available data already impose severe constraints on the hypothesis. In particular, the expected optical flux from the prior external shock model is higher than what the data allow. We conclude that if the shallow-decay X-ray component were from a prior emission component, it would have to be of an internal origin with optical flux suppressed.  
\end{abstract}

\begin{keywords}
gamma-rays: bursts, shock waves, radiation mechanisms: non-thermal
\end{keywords}

\section{Introduction}
Thanks to the vast collection of data on GRBs and their afterglows in the {\it Swift\/} era (Nousek et al. 2006; O'Brien et al. 2006; Willingale et al. 2007; Evans et al. 2009), we now have a detailed picture of the temporal behavior of the X-ray emission from these violent cosmic explosions.  Using light curves obtained by the X-Ray Telescope (XRT) onboard the {\it Swift\/} satellite, five phases that are common to many X-ray afterglow light curves were identified: a steep decay phase, a shallow decay phase, a normal decay phase, a post jet break phase, and X-ray flares (Zhang et al. 2006).  Not all bursts exhibit all five phases.  The normal decay and post jet break phases were predicted pre-{\it Swift\/} and are well-explained by the standard external shock afterglow model (M\'esz\'aros \& Rees 1997; Sari et al. 1998; Rhoads 1999; Sari et al. 1999).  {\it Swift\/} observations have challenged the GRB community to create physical models that account for the other three phases.  Of these phases, the shallow decay phase has definitely posed the biggest challenge.  Generally, no spectral evolution is observed across the temporal break from the shallow to the normal decay phase (Liang et al. 2007). This suggests that the shallow-to-normal transition should be either hydrodynamical or geometrical in origin.  Many such physical models of the shallow decay phase have been discussed in the literature, but none has been successful in interpreting all the observations (Zhang 2007 and references therein). 

An intriguing hypothesis was proposed by Yamazaki (2009). According to this hypothesis, there was already activity from the central engine before the GRB itself. The X-ray flux is already decaying before the GRB trigger. Due to a bad choice of reference time (the GRB trigger time), this power-law-decaying prior emission component (ejected at a time $-T_\Delta$ with respect to the GRB trigger time, which we define as $T=0$) appears as an artificial broken power law with a plateau stretching out to $T \approx T_\Delta$ in the log-log diagram.  In order to observe the single power law behavior of the prior emission component, one needs to move the time zero point to $-T_\Delta$, which we define as $t=0$.  According to Kobayashi \& Zhang (2007), the time zero point of external shock emission should coincide with the beginning of the corresponding central engine activity. So for prior emission it should be at $t=0$, while for the main GRB it should be at $T=0$\footnote{Quimby et al. (2006) have explored the effects that changing the afterglow reference time has on the X-ray and optical light curves for the ambiguous case of GRB 050319, a burst in our sample.  In our analysis of GRB 050319, we adopted their defined $t_{tr0}$ as $T=0$, which marks the start of the prompt emission even though the actual trigger happened later due to an instrumental slewing effect.}. The observed X-ray afterglow is dominated by the prior emission component, which outshines the X-ray emission associated with the main GRB outflow during the shallow and normal decay phases.  A systematic study of {\it Swift}/XRT light curves from GRBs detected before mid-2009 (Liang et al. 2009) suggests that the prior emission model provides a unified interpretation of both the canonical X-ray light curves and the light curves that seem to follow a single power law decay (which are the ones with negligibly short $T_\Delta$). In addition, Liang et al. (2009) also found that both types of light curves in their sample can be roughly explained by the external shock model.

If a prior explosion indeed happened, this prior outflow must sweep up the ambient medium and form an external shock. The long-lasting prior X-ray emission that is invoked to interpret the X-ray plateaus is most likely from this external shock. If this is the case, then the same external shock would give rise to an optical emission component that follows the same plateau behavior. This can be tested with the available optical data, especially during the earliest observational epoch. In general, optical emission observed during the prompt phase may be a mix of this prior emission component and several other emission components, including the optical counterpart of the prompt GRB emission, and the reverse shock and forward shock emission associated with the prompt outflow. As discussed in Ioka et al. (2006), the ejecta associated with the prompt emission would stream into the trail of the prior component blastwave with modified medium density profile, leading to another pair of forward and reverse shocks (for the dynamics setting of the three shock system, see Zhang \& M\'esz\'aros 2002). The relative importance of emission contribution to the optical band depends on many parameters. The reverse shock component can dominate or be outshone by the forward shock component in the optical band. The detected prompt optical emission flux is therefore allowed to be higher than the predicted flux. On the other hand, if the prompt optical flux or upper limit is already below the range of the predicted optical flux, then the external shock prior emission model would be ruled out. This is the motivation and strategy of this paper.

\section{Sample Selection and Data Reduction}

\begin{table*}
\setlength{\tabcolsep}{4pt}
\begin{minipage}{180mm}
\caption{Optical light curve references, redshifts, host galaxy extinction information, values of host galaxy and Galactic extinction, and spectral fit results for our sample GRBs.}
\begin{tabular}{llllllllll}
\hline
GRB & z & Host galaxy & Host & Galactic & Fitting & Galactic & Host & $\beta$ & $\tilde{\chi}^{2}~(\nu)$ \\
   &   & extinction & $A_{\lambda_{emit}}$ & $A_{\lambda_{obs}}$ & interval & $N_{H}*$ &  $N_{H}*$ &   &   \\
   &   & information & (mag) & (mag) & (ks) & ($10^{22}$ cm) & ($10^{22}$ cm) &   &   \\
\hline
050319$^{a}$ & 3.24$^{b}$ & $A_{V}=0.21\pm0.08$, MW$^{c}$ & $0.53\pm0.20$ & 0.028 & 6.24-82.77 & 0.0126 & 0.34 & $1.19\pm0.07$ & 0.83 (374)\\ 
050401$^{d}$ & 2.90$^{b}$ & $A_{V}=0.62\pm0.06$, SMC$^{e}$ & $2.4\pm0.2$ & 0.167 & 0.46-20.09 & 0.044 & 1.57 & $1.00\pm0.07$ & 0.78 (365)\\
050822 & 1.2$^{f}$ & unknown & ? & - & 6.41-516.15 & 0.014 & 0.118 & $0.90\pm0.04$ & 1.03 (421)\\
051109A$^{g}$ & 2.35$^{b}$ & $A_{V}<0.10$, SMC$^{c}$ & $<0.3189\pm0.0008$ & 0.487 & 3.73-636.40 & 0.16 & 0.93 & $1.29\pm0.03$ & 1.03 (560)\\
060729 & 0.54$^{b}$ & $A_{V}=0.04\pm0.02$, LMC$^{c}$ & $0.05\pm0.03$ & - & 0.70-2,221.24 & 0.045 & 0.099 & $1.14\pm0.01$ & 1.16 (749)\\
060927$^{h}$ & 5.47$^{h}$ & $E_{B-V}=0.07\pm0.03^{h}$, SMC & $1.1\pm0.5$ & - & 0.11-10.86 & 0.046 & 0.860 & $1.0\pm0.1$ & 0.50 (208)\\
061121$^{i}$ & 1.31$^{b}$ &  $A_{V}=0.40\pm0.04$, LMC$^{c}$ & $0.80\pm0.08$ & 0.116 & 0.59-348.63 & 0.04 & 0.607 & $1.09\pm0.03$ & 0.88 (587)\\
061222A & 2.088$^{j}$ & $A_{V}>5.0^{j}$, SMC & $>14.74\pm0.04$ & - & 0.23-723.42 & 0.09 & 3.82 & $1.13\pm0.03$ & 1.15 (692)\\
\hline
\end{tabular}

*Taken from Evans et al. (2009) and held constant in spectral fit; $^{a}$Optical light curve data from Greco et al. (2005a), Kiziloglu et al. (2005), Misra et al. (2005a), Sharapov et al. (2005), Wo\'zniak et al. (2005), Quimby et al. (2006), Huang et al. (2007), and Kamble et al. (2007); $^{b}$Liang et al. (2007); $^{c}$Schady et al. (2010); $^{d}$Optical light curve data from D'Avanzo et al. (2005), Greco et al. (2005b), Kahharov et al. (2005), Misra et al. (2005b), Rykoff et al. (2005), De Pasquale et al. (2006), Watson et al. (2006), and Ghirlanda et al. (2007); $^{e}$Watson et al. (2006); $^{f}$Butler (2007); $^{g}$Optical light curve data from Yost et al. (2007a); $^{h}$Optical light curve data, redshift, and host galaxy extinction information from Ruiz-Velasco et al. (2007); $^{i}$Optical light curve data from Efimov et al. (2006), Halpern et al. (2006), Halpern \& Armstrong (2006a,b), Melandri et al. (2006), Page et al. (2007), and Uehara et al. (2011); $^{j}$Perley et al. (2009)
\end{minipage}
\end{table*}

In order to create a sample for this study, we need bursts with both X-ray plateaus (which can be fit within the prior emission model) and early optical observations (either detections or upper limits). We searched through the literature to find the GRBs which either clearly exhibit a shallow decay segment in their XRT light curve or have been fit with a shallow decay segment.  Specifically, the candidate bursts were taken from Table 1 of Liang et al. (2007) and Table 2 of Liang et al. (2009).  We further narrowed down the sample to those GRBs with early optical observations, using Table 5 from Yost et al. (2007a) and Table 2 from Yost et al. (2007b) for detections, and Table 3 from Yost et al. (2007b) for upper limits. Thus, the effective cutoff date for our sample is December 2006, about two years into the {\it Swift\/} mission. Altogether, our sample consists of eight bursts (Table 1) with a total of 11 early optical detections and 17 early optical upper limits. In the interest of comparing the optical light curves of these GRBs to those predicted by the external shock prior emission model, we have included as much of the available optical data for these bursts as possible.  The references for the optical light curve data are given in Table 1.  Since most of the prompt optical observations from Yost et al. (2007a,b) are in the {\it R}-band, we include only the {\it R}-band optical light curves. The only exception is GRB 060927, for which the included prompt optical observations and optical light curve data are at a wavelength ($8190~\mathring{A}$) near the {\it i}-band (Ruiz-Velasco et al. 2007).

Before we can compare the prompt optical observations and optical light curves from the literature to the predictions of the external shock prior emission model, the optical data must be corrected for Galactic and host galaxy extinction.  We corrected the optical data for Galactic extinction using the empirical Milky Way (MW) extinction law, extinction coefficients, and $R_{V}=3.08$ from Pei (1992), along with the values of $A_B$ from Schlegel et al. (1998).  In order to correct the optical data for host galaxy extinction, we searched the literature on each burst in our sample for the extinction law and $A_V$ or $E_{B-V}$ value which provided the best fit to its afterglow spectral energy distribution (SED).  We then used the appropriate results from the literature to calculate the host galaxy extinction, once again relying on the empirical MW, Large Magellanic Cloud (LMC), and Small Magellanic Cloud (SMC) extinction laws and parameters from Pei (1992).

The host galaxy extinction information and references for each GRB are given in Table 1.  For GRB 060927 and GRB 061222A, the cited authors were unable to distinguish between the three extinction laws.  We chose the SMC extinction law based on the conclusions of Schady et al. (2007, 2010), that, in most cases, it provides an acceptable fit to the host galaxy extinction profile.  It should also be noted that the host galaxy extinction of GRB 050401 is the subject of debate.  We have decided to use the analysis of Watson et al. (2006), although higher values of $A_V$ have been advocated by De Pasquale et al. (2006) and Kamble et al. (2009).

The calculated values (in magnitudes) of the host galaxy extinction $A_{\lambda_{emit}}$ and Galactic extinction $A_{\lambda_{obs}}$ are given in Table 1.  The error on $A_{\lambda_{emit}}$ is due to propagating the error on the host galaxy value of $A_V$ or $E_{B-V}$ taken from the literature, and the error on $R_V$ given by Pei (1992).  Since the host galaxy $A_V$ values for GRB 051109A and GRB 061222A are limits, the errors on their $A_{\lambda_{emit}}$ only take into account the error on $R_V$.  We were unable to calculate the host galaxy extinction for GRB 050822 since there is no information about its host galaxy extinction profile in the literature.  We did not calculate Galactic extinction for GRB 050822, GRB 060729, GRB 060927, and GRB 061222A since all of the included optical flux densities for these bursts have already been corrected for Galactic extinction by either Yost et al. (2007a,b) or Ruiz-Velasco et al. (2007). The errors on  $A_{\lambda_{obs}}$ were negligible.

The {\it Swift}/XRT data for the GRBs in our sample are taken from the {\em Swift} data archive\footnote{\url{ftp://legacy.gsfc.nasa.gov/swift/data/}}. We developed an IDL package which invokes the standard HEASoft tools (e.g., Xselect, Ximage, Xspec, Swift tools) to automatically process the XRT data for any given burst. The details of the procedure of our package are described in Zhang et al. (2007b). Our final data products\footnote{Available from \url{http://grb.physics.unlv.edu/swift}} include the XRT light curves (in physical units), extracted over the energy range $0.3-10~keV$.  These data products have been applied to study statistical properties of X-ray afterglows in a series of papers (Zhang et al. 2007b; Liang et al. 2007, 2008, 2009).

 For the bursts in our sample, the spectral index $\beta$ for the XRT band is obtained by fitting the time-integrated spectrum (Evans et al. 2009) of the shallow and normal decay phases with a simple power law model in Xspec. The spectral fit also includes two-component neutral hydrogen column density $N_{H,host}$ and $N_{H,Gal}$, with corresponding values taken  from the late-time spectral data in the Swift/XRT GRB spectrum repository\footnote{\url{http://www.swift.ac.uk/xrt_spectra/}} (Evans et al. 2009). Their values were held constant in our spectral fit, which are collected in Table 1 along with the GRB redshift $z$ and the spectral fitting results.

All the errors quoted in this paper correspond to the $1\sigma$ errors on the quantities, unless otherwise indicated.  The convention $F_{\nu}\varpropto{ }t^{-\alpha}\nu^{-\beta}$ is adopted.  The zero points and effective frequencies for the optical bands from Bessell (1979) are adopted.

\section{Model}

\begin{table*}
\begin{minipage}{180mm}
\caption{Results of temporal fits and spectral regime/model determinations.}
\begin{tabular}{lllllllll}
\hline
GRB &  Fitting & $F_0$  & $T_\Delta$ & $\alpha$ & $\tilde{\chi}^{2}~(\nu)$ & Regime & Model & $\alpha_{pred}$ \\
   & interval &  ($10^{-11}$ erg & (ks) &   &   &   &   &   \\
   & (ks) & \rm{s$^{-1}$~cm$^{2}$})  &   &   &   &   &   &   \\
\hline
050319 & 6.11-84.79 & $2.1\pm0.3$ & $17\pm10$ & $1.5\pm0.4$ & 0.51 (27) & $\nu_X>\nu_c$ & ISM, $p>2$ & $1.29\pm0.11$ \\ 
050401 &  0.14-20.09 & $87\pm4$ & $0.53\pm0.05$ & $1.13\pm0.03$ & 2.21 (132) & $\nu_X>\nu_c$ & ISM, $p>2$* & $1.00\pm0.11$ \\
050822 & 6.41-523.32 & $1.5\pm0.3$ & $12\pm5$ & $1.1\pm0.1$ & 1.24 (72) & $\nu_m<\nu_X<\nu_c$ & ISM, $p>2$* & $1.35\pm0.06$ \\
051109A & 3.73-639.16 & $15\pm2$ & $3.3\pm0.7$ & $1.33\pm0.04$ & 1.17 (94) & $\nu_X>\nu_c$ & ISM, $p>2$* & $1.44\pm0.05$ \\
060729 & 0.70-2,221.24 & $2.93\pm0.05$ & $120\pm7$ & $1.96\pm0.05$ & 1.69 (813) & $\nu_m<\nu_X<\nu_c$ & ISM, $p>2$* & $1.71\pm0.02$ \\
060927 & 0.11-10.86 & $4.4\pm0.5$ & $2.7\pm1.8$ & $2.5\pm1.0$ & 1.19 (29) & $\nu_m<\nu_X<\nu_c$ & Wind, $p>2$ & $2.0\pm0.2$ \\
061121 & 0.30-353.10 & $35.8\pm0.9$ & $2.6\pm0.1$ & $1.35\pm0.02$ & 1.18 (412) & $\nu_X>\nu_c$ & ISM, $p>2$* & $1.14\pm0.05$ \\
061222A & 0.23-724.62 & $10.2\pm0.2$ & $8.5\pm0.4$ & $1.56\pm0.02$ & 1.72 (317) & $\nu_m<\nu_X<\nu_c$ & ISM, $p>2$* & $1.70\pm0.05$ \\
\hline
\end{tabular}

*$\alpha$ and $\alpha_{pred}$ not consistent within error for this regime and model
\end{minipage}
\end{table*}

In this paper we focus on the XRT data in the shallow and normal decay phases (dark black points in Fig. 3).  We do not consider the post jet break phase as the jet break happens much later. Using the prior emission model and Eq. [\ref{X2}] from Liang et al. (2009) as our prescription, we fit\footnote{Using the routine MPFITFUN.PRO (Markwardt 2009): \url{http://purl.com/net/mpfit}} the shallow and normal decay segments of each XRT light curve to the function
\begin{equation}
F=F_0 \left(\frac{T+T_\Delta}{T_\Delta}\right)^{-\alpha} \label{X1}
\end{equation}
allowing the normalization constant $F_0$, time shift constant $T_\Delta$, and temporal decay index $\alpha$ to vary in the fit. Here $T$ is the time measured since the GRB trigger. The fitting intervals are loosely based on the time ranges given in Table 1 of Liang et al. (2007).  Once the parameters have been determined for a particular burst, Eq. [\ref{X1}] can be used to extrapolate the X-ray flux due to the prior emission component from the shallow decay segment to earlier times (i.e., to the time of the early optical observations of interest).  The results of the XRT light curve fits are given in Table 2.

The external shock model predicts a broken power law synchrotron spectrum (M\'esz\'aros \& Rees 1997; Sari et al. 1998).  For a constant density medium, a fast cooling spectrum ($\nu_c < \nu_m$) is expected at early times, which transitions to a slow cooling spectrum ($\nu_m < \nu_c$) at later times (Sari et al. 1998).  In our analysis, since $T_\Delta\sim10^3-10^4~s$ (typical observed shallow-to-normal break time), the prior emission external shock should have already entered the slow cooling regime by the time of the GRB trigger. Thus, we apply slow cooling throughout the modeling.

We first determine the spectral regime in which the XRT band is located by using the X-ray $\alpha$ and $\beta$ data and comparing the data with the closure relations of the external shock models (Zhang \& M\'esz\'aros 2004).  Our approach is similar to that used by Racusin et al. (2009).  We use the value of $\beta$ from our spectral fit to evaluate the slow cooling closure relations, with each model/regime combination yielding a predicted value of the temporal decay index $\alpha_{pred}$.  We then compare the values of $\alpha_{pred}$ to the value of $\alpha$ obtained from our temporal fit, looking for the model/regime combination for which $\alpha-\alpha_{pred}=0$ (within $1\sigma$ error).  Of course,  the decision typically was not clear cut.  For six out of the eight GRBs in our sample, $\alpha$ was not consistent with $\alpha_{pred}$ from any of the model/regime combinations within error.  In these cases, we chose the model and regime with which the data was most consistent (minimizing $|\alpha-\alpha_{pred}|$).

Ultimately, the XRT band is found to be either in the $\nu_m<\nu_X<\nu_c$ or the $\nu_X>\nu_c$ spectral regime\footnote{In some cases, a burst is consistent with both ISM and Wind models within the same spectral regime. For the analysis below, the medium density profile does not enter the problem. So we care about the spectral regime only.}. The results of our spectral regime and model determinations are given in Table 2.  One ambiguous case was GRB 050822. Its XRT data were nearly equally consistent with three model/regime combinations: $\nu_X>\nu_c$ for ISM and Wind ($p<2$), and $\nu_m<\nu_X<\nu_c$ for ISM ($p>2$).  We chose the last model/regime combination for GRB 050822 because the resulting predicted range for the optical flux actually encompasses the predicted ranges corresponding to the other two model/regime choices.

Next, we apply the XRT light curve fit results to extrapolate the prior emission X-ray flux back to the time of the optical observations ($T_{obs}$), and derive the mean X-ray spectral density as
\begin{equation}
F_X(T_{obs})=\frac{F_0}{\nu_X}\left(\frac{T_{obs}+T_\Delta}{T_\Delta}\right)^{-\alpha} \label{X2}
\end{equation}
where $\nu_X \approx 4.8 \times 10^{17}$ Hz (2 keV)\footnote{Approximate log average of the XRT energy range} is applied.

\begin{figure}
\begin{tabular}{l}
\includegraphics*[bb=142 370 465 770,scale=0.7]{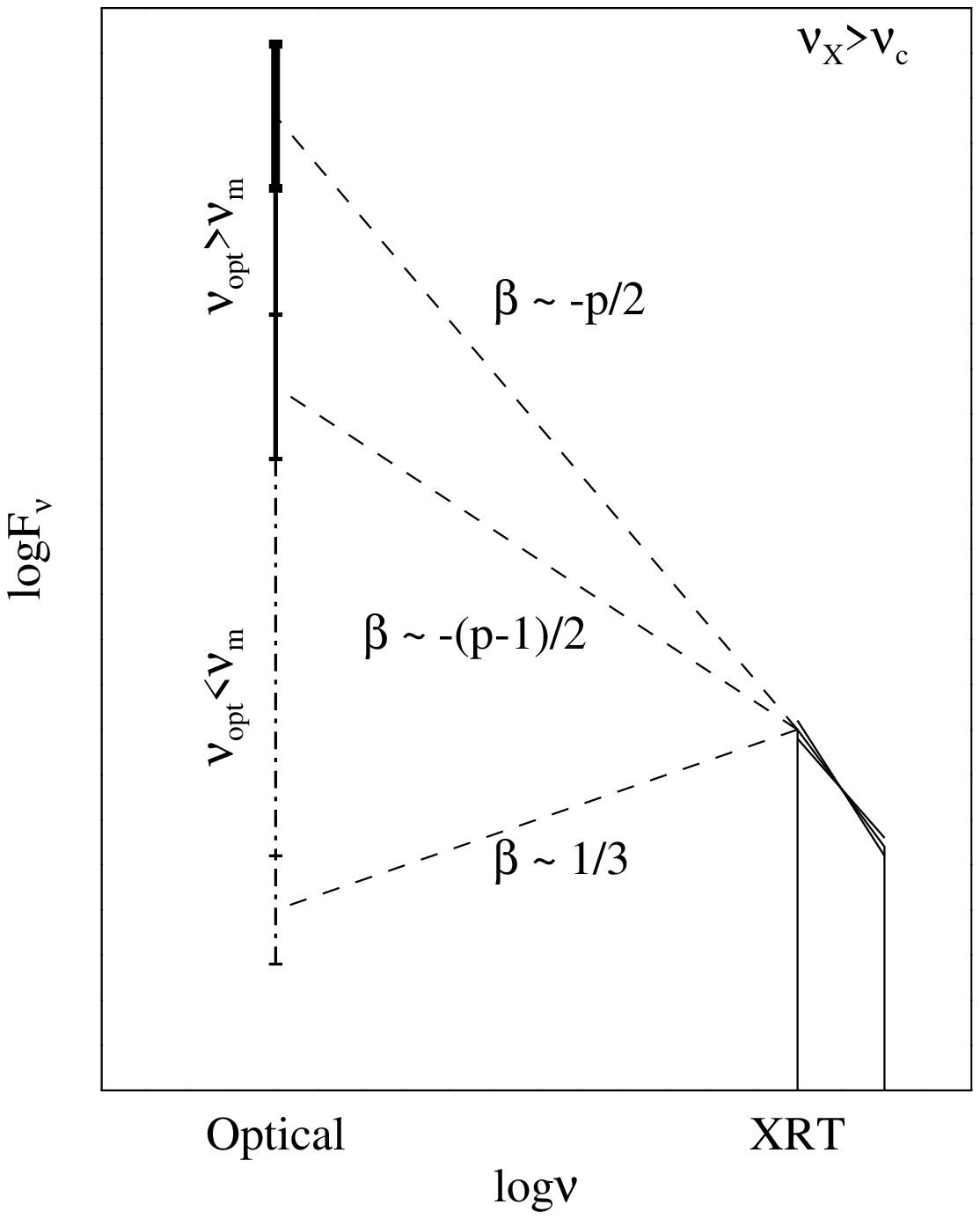}\\
\includegraphics*[bb=142 370 465 770,scale=0.7]{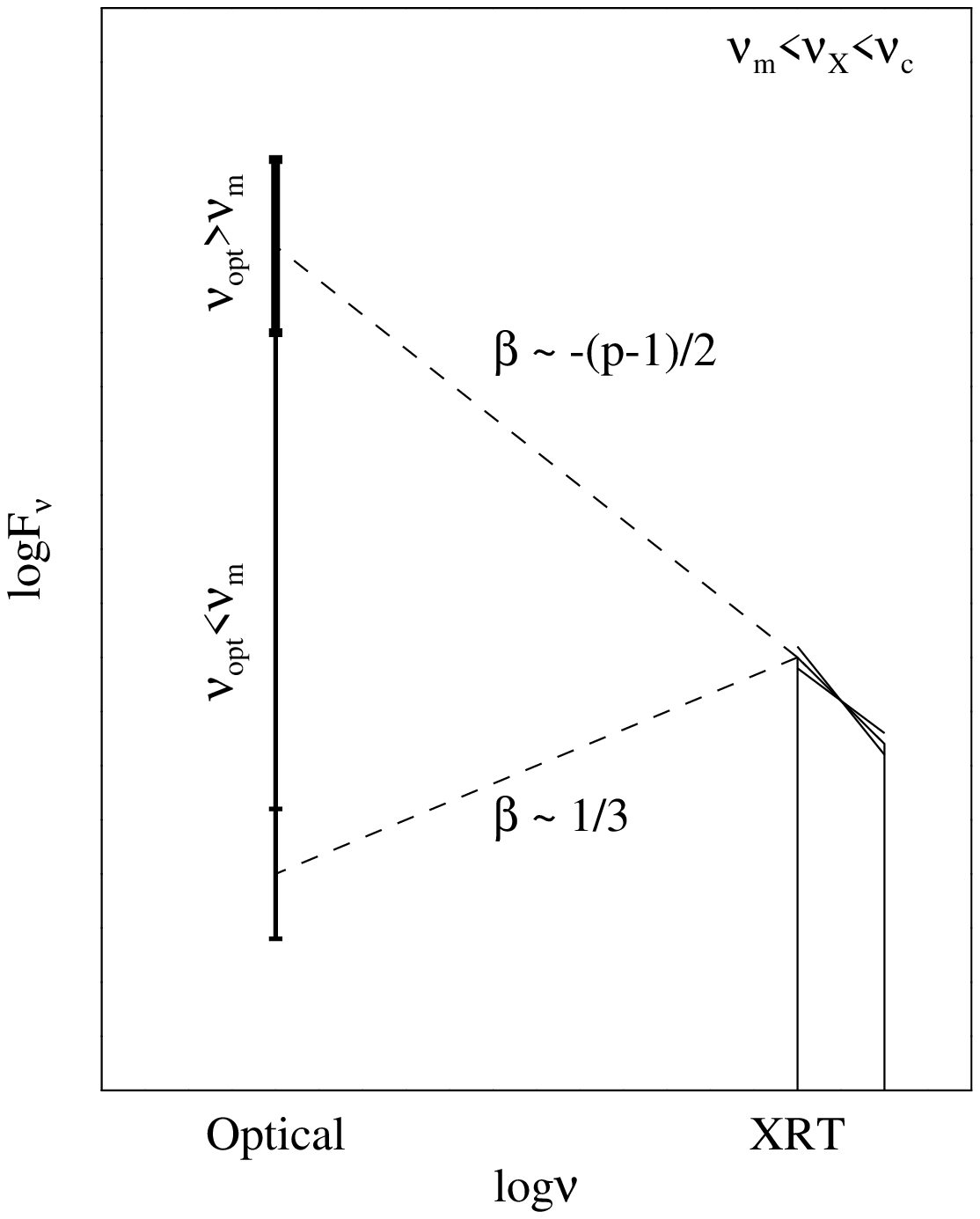}\\
\end{tabular}
\caption{Cartoon spectra based on the results of Sari et al. (1998).  For each possible spectral regime for $\nu_{X}$, there are minimum and maximum possibilities for the corresponding optical flux density (section 3).}
\end{figure}

We then apply the slow cooling model (Fig. 1b of Sari et al. 1998) to predict the range of optical flux density. In the case that $\nu_X>\nu_c$ (Fig. 1a), the maximum possible optical flux density occurs when $\nu_{opt} > \nu_c$, so that the optical band lies in the same spectral regime and one can do a simple extrapolation of X-ray flux density based on the power law spectral fits (top dashed line):
\begin{equation}
F_{opt,max}(T_{obs})=F_X(T_{obs})\left(\frac{\nu_{opt}}{\nu_X}\right)^{-\beta} 
\label{X3}
\end{equation}
If $\nu_{opt}$ is below $\nu_c$, then most likely $\nu_m<\nu_{opt}<\nu_c$. The minimum possible optical flux density in this regime occurs when $\nu_X$ is just above $\nu_c$, so that
\begin{equation}
F_{opt,min}(T_{obs})=F_X(T_{obs})\left(\frac{\nu_{opt}}{\nu_X}\right)^{-(\beta-{1}/{2})} 
\label{X4}
\end{equation}
This is marked as the second dashed line in Fig. 1a. In the most extreme cases, one may have both $\nu_m$ and $\nu_c$ between the optical and X-ray bands. Therefore the lowest flux density allowed corresponds to $\nu_m = \nu_c \lesssim \nu_X$, so that
\begin{equation}
F_{opt,min}(T_{obs})=F_X(T_{obs})\left(\frac{\nu_{opt}}{\nu_X}\right)^{{1}/{3}} 
\label{X5}
\end{equation}
This corresponds to the third dashed line in Fig. 1a. 

In the case that $\nu_m<\nu_X<\nu_c$ (Fig. 1b), the maximum possible optical flux density occurs when $\nu_X$ is just below $\nu_c$ and $\nu_{opt}$ is in the same regime as $\nu_X$ (Eq. [\ref{X3}]).  The minimum possible optical flux occurs when $\nu_X$ is just above $\nu_m$ and $\nu_{opt}<\nu_m \lesssim \nu_X$ (Eq. [\ref{X5}]). 

In both Fig. 1a and Fig. 1b, the thicker part of the range corresponds to the maximum prediction, for which there are no spectral breaks between the X-ray and optical bands.  The thinner part of the range, which includes the minimum prediction, represents the case in which there is one spectral break between the bands.  The dash-dotted line in Fig. 1a corresponds to the extreme case in which there are two spectral breaks between the X-ray and optical bands.  In all cases, we have introduced an error on the prediction, which includes the uncertainties in $\beta$, $F_0$, $T_\Delta$, and $\alpha$, with the prior emission model and external shock synchrotron model applied and errors properly propagated.

The final predicted range for the optical flux density can then be compared to the optical observations with proper extinction correction. If all of the extinction-corrected prompt optical observations fall within their predicted ranges, then the data would show no inconsistency with the hypothesis of an external-shock-origin prior emission.  If some or all of the observations lie above their predicted ranges, this would not count as evidence against this hypothesis, because the observed optical flux can have contributions from the internal dissipation region and the reverse/forward shock associated with the prompt emission, besides the contribution from the prior emission component. However, an observation that lies below its predicted range would put a severe constraint on the external shock prior emission model.

\begin{figure}
\includegraphics*[bb=75 0 525 285,scale=0.55]{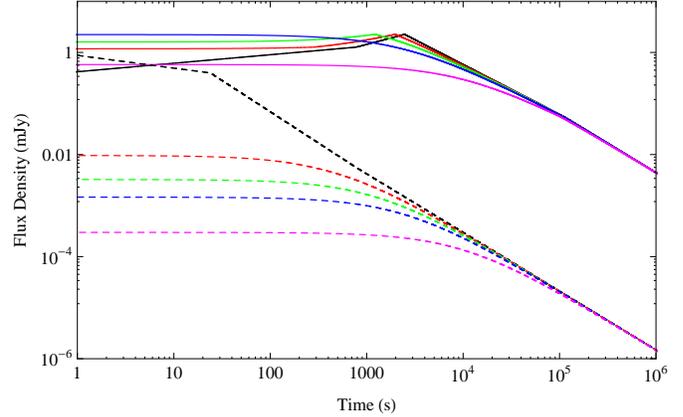}\\
\caption{Example optical (solid) and X-ray (dashed) light curves derived from the external shock prior emission model with typical parameter values: $t_{m}=2500{\rm~s}$ ($p=2.3$, $\epsilon_{e}=0.3$, $\epsilon_{B}=0.01$, $E_{K}=10^{52} {\rm~erg}$, $n=0.5$, $z=1.0$) and  $T_\Delta=0{\rm~s}$ (black), $500{\rm~s}$ (red), $1250{\rm~s}$ (green), $2500{\rm~s}$ (blue), $10000{\rm~s}$ (pink).  These light curves were generated by our GRB afterglow code which uses the results of Sari et al. (1998) and Zhang et al. (2007a) as a prescription.}
\end{figure}

The interpretation becomes more complicated in the cases where later optical observations are available.  In these cases, one may examine the overall shape of the optical light curve to determine whether or not it is consistent with the external shock prior emission model. If the prior emission time is chosen as the zero point, the X-ray lightcurve and optical lightcurve typically have different shapes. In particular, if $\nu_{opt} < \nu_m$ early on, the optical lightcurve would display an initial rise with $F_{opt} \propto t^{1/2}$ (where $t$ is time measured from the prior emission zero point) followed by a decay (Sari et al. 1998). With $T_\Delta$ shift, the shapes of both X-ray and optical lightcurves are distorted. In particular, the optical lightcurve as measured since the GRB trigger may be well fit by a smooth broken power law
\begin{equation}
F=F_0\left[\left(\frac{T+T_\Delta}{T_m+T_\Delta}\right)^{-\frac{w}{2}}+\left(\frac{T+T_\Delta}{T_m+T_\Delta}\right)^{w\alpha}\right]^{-\frac{1}{w}}
\label{X6}
\end{equation}
where $T_m$ is the time referenced from the GRB trigger when $\nu_m$ crosses $\nu_{opt}$ and $w$ is the smoothing factor (Liang et al. 2007).  Figure 2 shows the comparison between the X-ray and optical lightcurves for various values of $T_\Delta$. It is found that the difference between the shapes of the two lightcurves reduces as $T_\Delta$ increases, and essentially disappears as $T_\Delta$ exceeds $t_m$ (the time referenced from prior emission when $\nu_m$ crosses $\nu_{opt}$).

If the optical lightcurve does not follow the predicted shape of the external shock prior emission model, one would conclude that it is dominated by another component. In that case, the allowed optical emission level from the external shock prior emission component should be at most the highest predicted prior emission lightcurve allowed by the data.  Additionally, the predicted prior emission optical light curve should lie within the predicted range delineated by the bars in Figure 1.

\section{Results and Discussion}

\begin{figure*}
\begin{tabular}{l @{} l @{} l}
\includegraphics[bb=108 368 476 777,scale=.47]{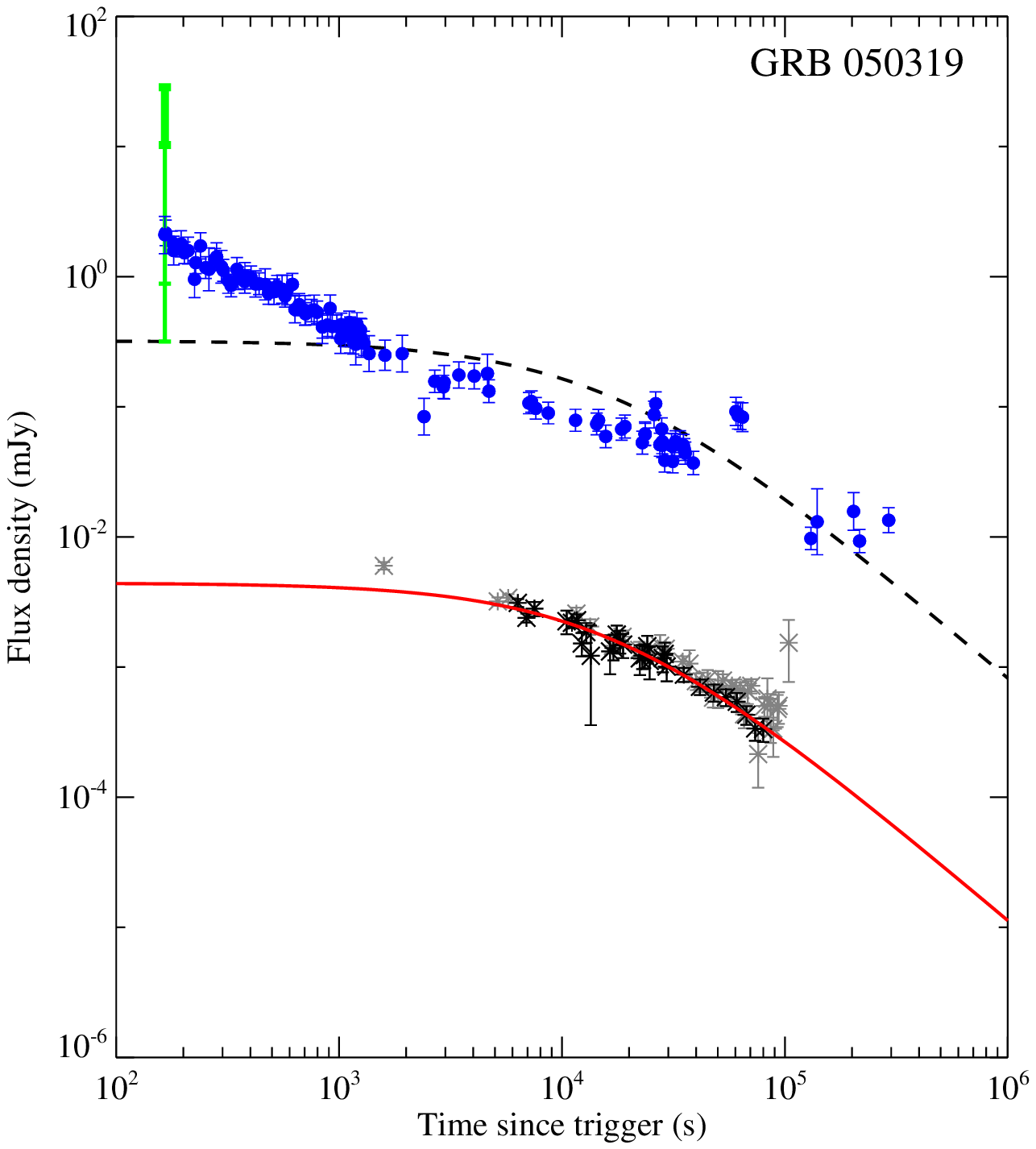} & 
\includegraphics[bb=108 368 476 777,scale=.47]{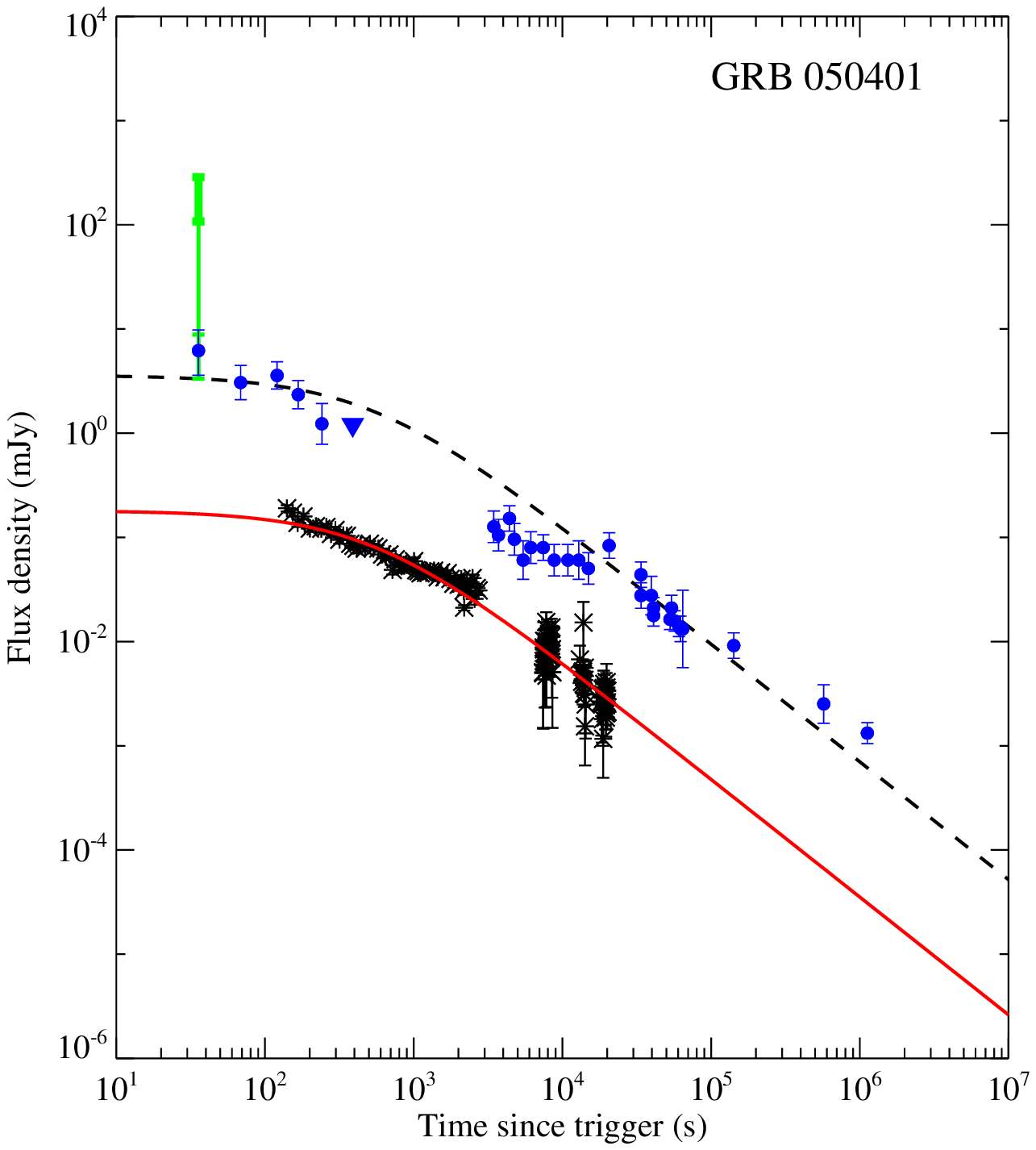} & 
\includegraphics[bb=108 368 476 777,scale=.47]{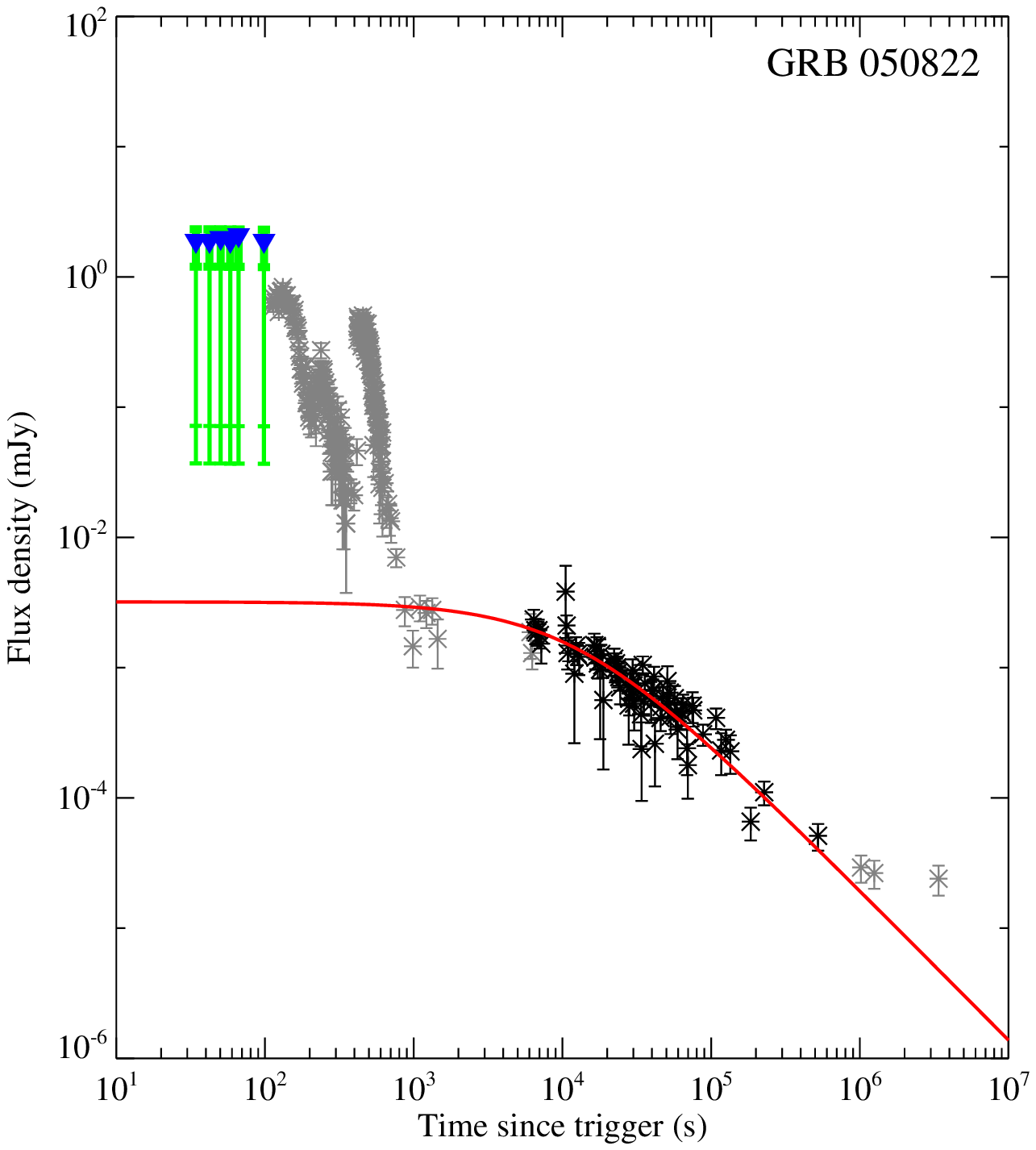}\\
\includegraphics[bb=108 368 476 777,scale=.47]{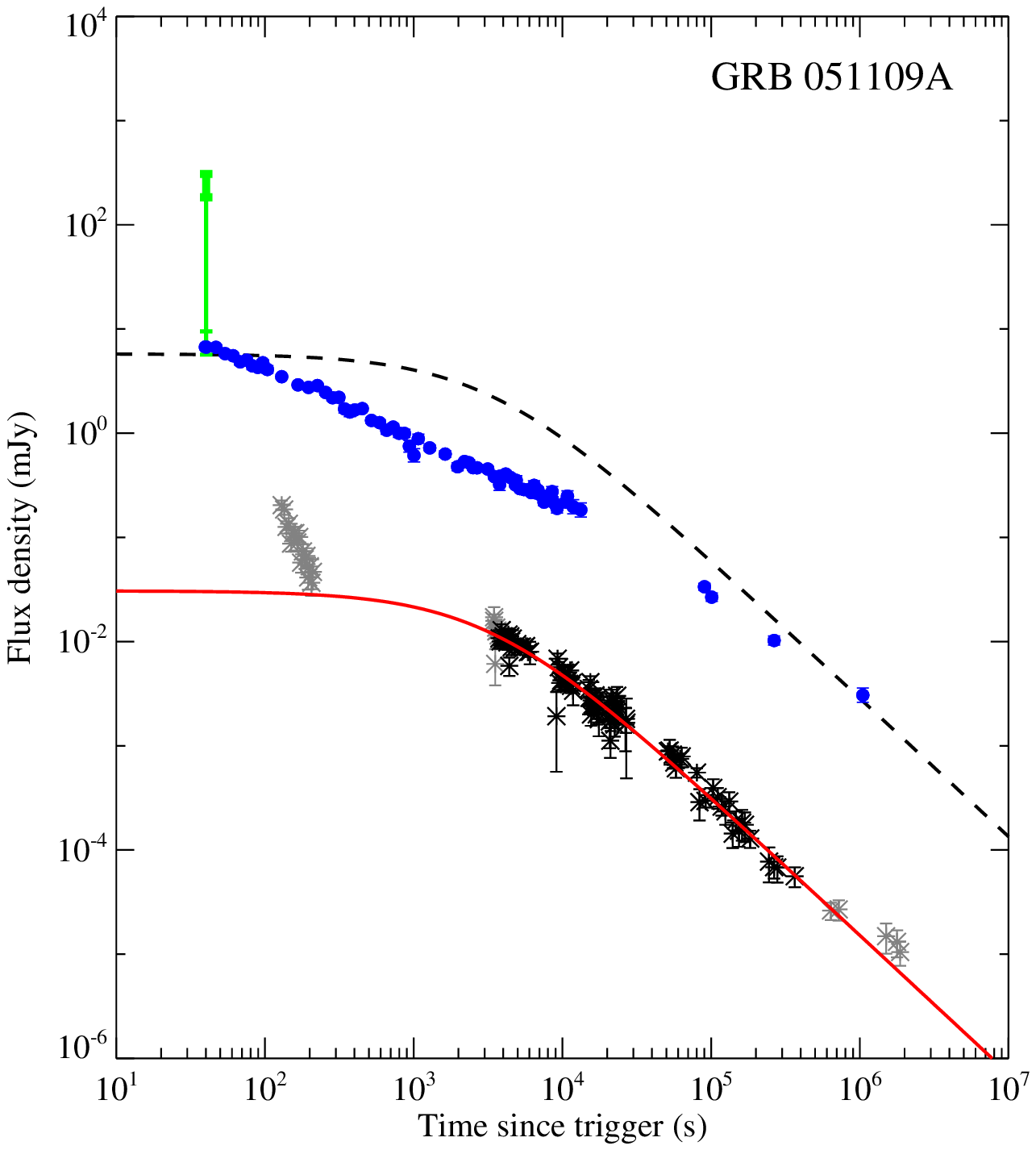} &
\includegraphics[bb=108 368 476 777,scale=.47]{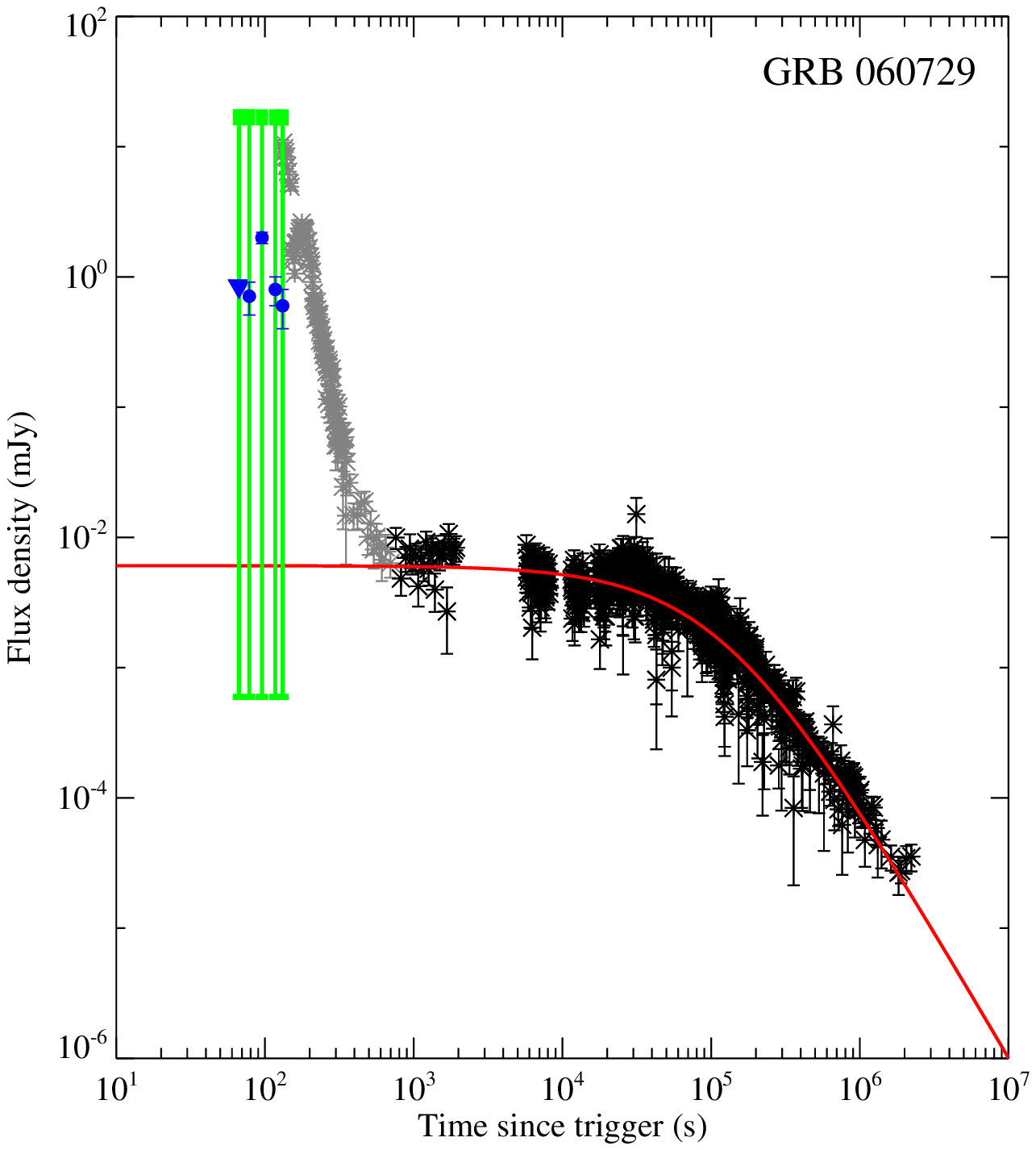} &
\includegraphics[bb=108 368 476 777,scale=.47]{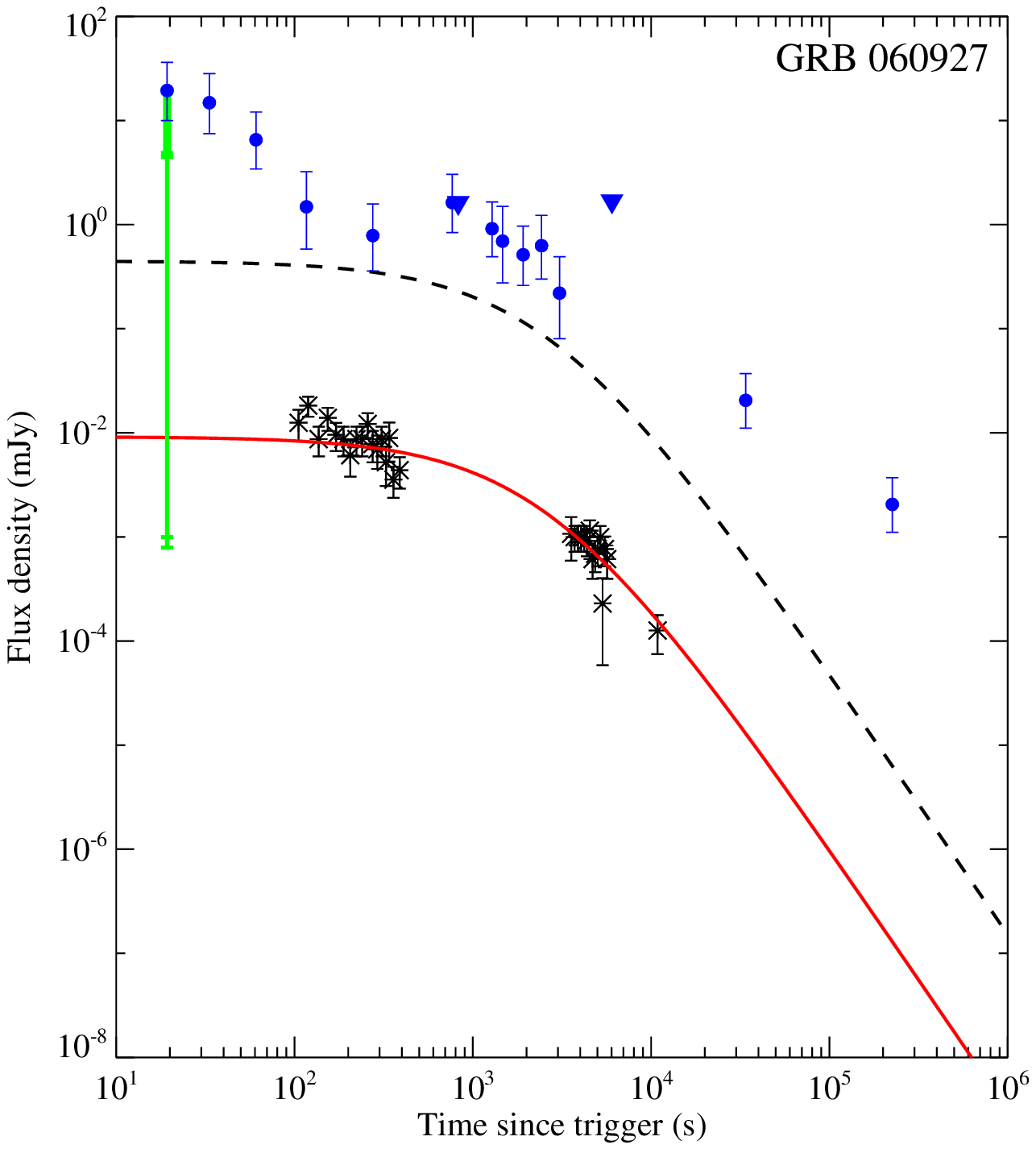}\\
\includegraphics[bb=108 368 476 777,scale=.47]{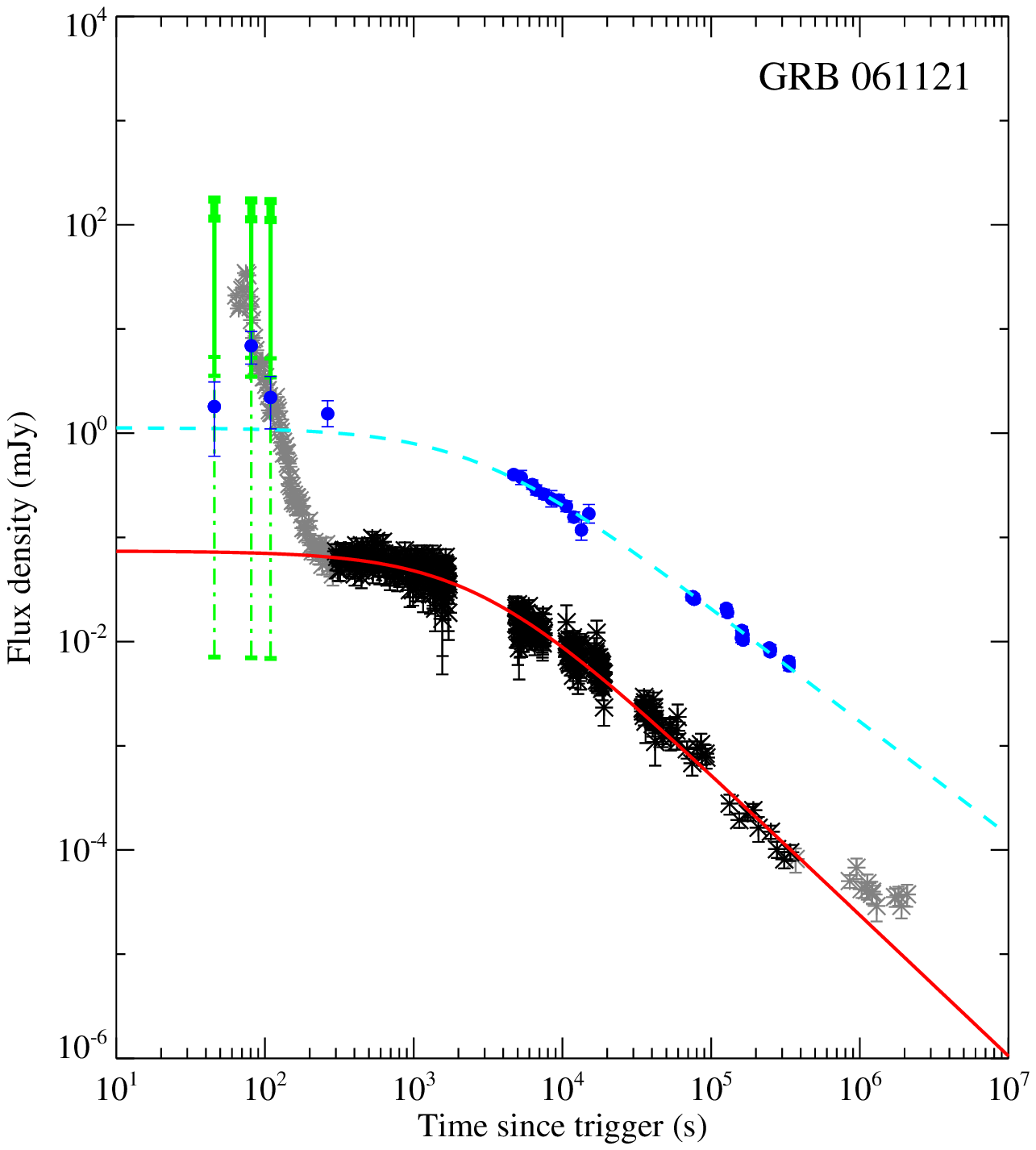} &
\includegraphics[bb=108 368 476 777,scale=.47]{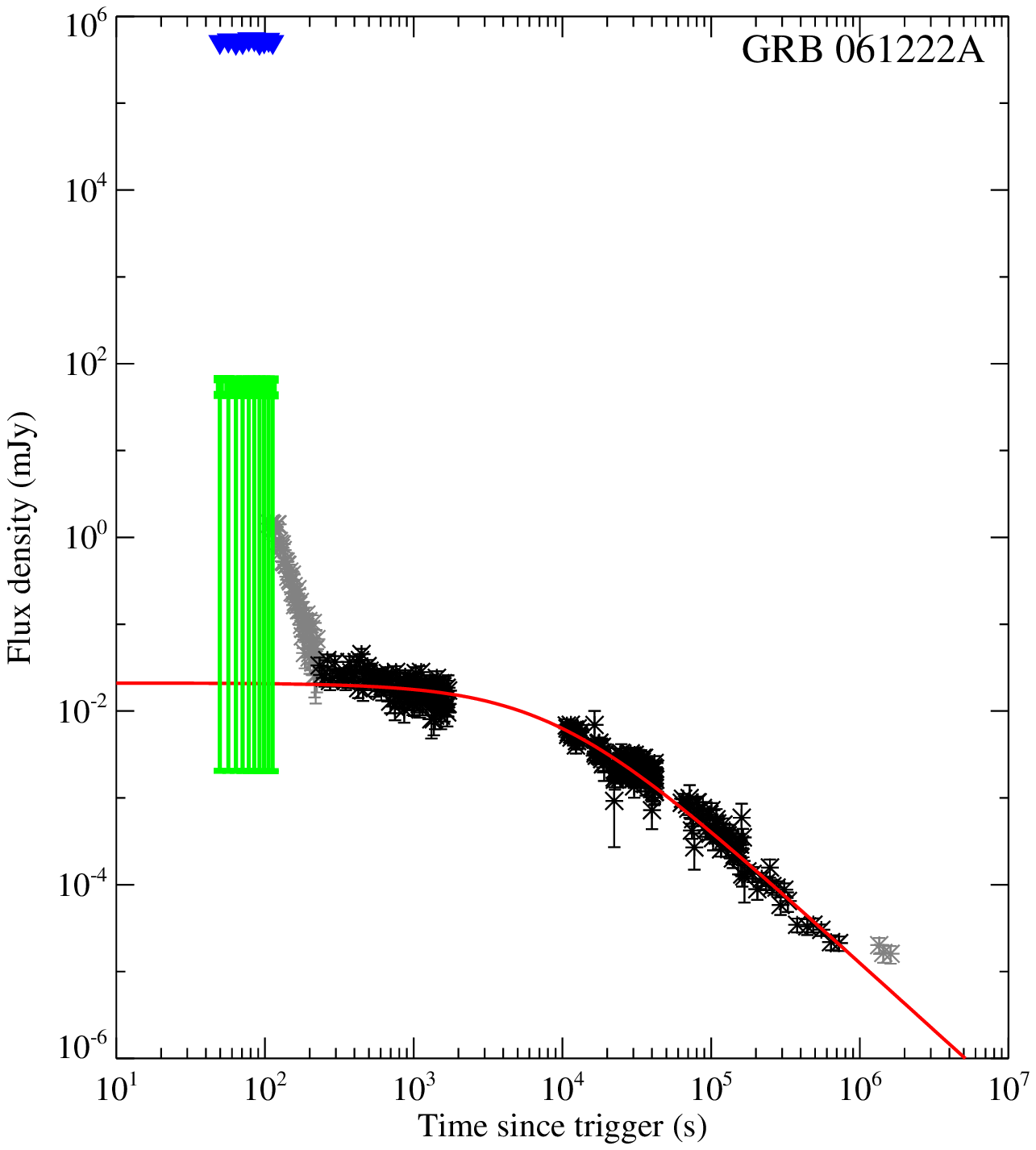} & \\
\end{tabular}
\caption{The grey and black asterisks are the XRT light curve data.  The red line is the fit of Eq. [\ref{X1}] to the black XRT light curve data points. The green bars represent the predicted ranges for the optical flux density at $T_{obs}$.  The thicker part of the green bar corresponds to the maximum prediction for which there are no spectral breaks between the XRT and optical bands.  The thinner part of the green bar, which includes the minimum prediction, is obtained by assuming there is a spectral break between the bands.  For GRB 061121, the dash-dotted portion of the green bar represents the extreme minimum prediction obtained by assuming there are two spectral breaks between the XRT and optical bands.  The dark blue circles and triangles represent the optical detections and upper limits, respectively.  The dashed black lines are the predicted optical light curves from the external shock prior emission model, as constrained by the green ranges and the optical data.  For GRB 061121, the dashed light blue line is the fit of Eq. [\ref{X1}]  to the optical light curve data in which $T_\Delta$ is fixed to the value from the XRT light curve fit.}
\end{figure*}

The results of the spectral fits, burst redshifts, extinction information, and the inclusive list of the references for the optical light curve data are given in Table 1.  The XRT light curve fits and spectral regime determinations are given in Table 2. The XRT light curve data, associated fits, extinction-corrected optical observations, and predicted ranges are shown in Figure 3. The green bars in the plots have the same meanings as the bars shown in the cartoon spectra in Figure 1 and explained in section 3. The dashed black lines are the maximum predicted optical light curves from the external shock prior emission model allowed by the data.  If the optical data constrain the dashed black line to be below the green range, for self-consistency we instead plot a dashed line corresponding to the minimum optical light curve allowed by the external shock prior emission model, even though it clearly contradicts the data.  The dashed light blue line in the plot for GRB 061121 is a fit to its optical light curve data (discussed in section 4.2). Looking at Figure 3, one can see that the results for each GRB are a bit different.  In general, we can assign the bursts to one of two groups: those that are consistent (or at least, not inconsistent) with the external shock prior emission hypothesis, and those that are inconsistent with the external shock prior emission hypothesis.

\subsection{``Consistent" Bursts: GRB 050822, GRB 060729, GRB 060927, and GRB 061222A}
  
The early optical upper limits of GRB 050822 lie within the thick part of the green range, corresponding to the maximum prediction in which $\nu_m<\nu_{opt}<\nu_X<\nu_c$.  However, since these data are just upper limits and not actual detections, the actual level of optical flux from this dark burst could lie anywhere in the green range or even below the green range.  In fact, since we were unable to correct these upper limits for host galaxy extinction, the limits (and thus the actual level of optical flux) could even be above the green range.  Regardless, the facts that these are only upper limits, and that even without correction for host galaxy extinction they lie within the green range, lead us to the conclusion that GRB 050822 is not inconsistent with the external shock prior emission hypothesis.

GRB 060729 has been studied (e.g., by Grupe et al. 2007, Curran et al. 2009) in many optical bands. However, there is no additional {\it R}-band data in the literature.  We find the early {\it R}-band optical upper limit and detections of GRB 060729, which have been corrected for both Galactic and host galaxy extinction, lie within the thinner part of the green range.  This is consistent with the case in which there is one spectral break between the X-ray and optical bands:  $\nu_{opt}<\nu_m<\nu_X<\nu_c$.  Due to the lack of additional {\it R}-band data, we cannot make any further comparisons of this burst to predictions of the external shock prior emission hypothesis, and conclude that the available {\it R}-band data for GRB 060729 are not inconsistent with this hypothesis.

For GRB 060927, the early optical data is at a wavelength near the {\bf {\it i}-band}, and fortunately there is additional optical data in this band in the literature.  The prompt {\bf {\it i}-band} optical flux density falls within the thicker part of the green predicted range, corresponding to the maximum prediction in which  $\nu_m<\nu_{opt}<\nu_X<\nu_c$.  It is clear that the overall shape of the optical light curve is not consistent with the smooth, $T_\Delta$-shifted single power law predicted by the external shock prior emission model.  This suggests that, if there is a prior emission external shock optical component, it is dominated by other optical components.  The dashed black line shows the maximum prior emission optical light curve allowed by the data.

The results for GRB 061222A are similar to those for GRB 050822, except in this case, the early optical upper limits are nearly four orders of magnitude above the green range.  This is explained by the large amount of host galaxy extinction (nearly 15 magnitudes) for this burst.  Because GRB 061222A is a dark burst, we don't know whether the actual level of optical flux is consistent with the prediction of the external shock prior emission model.  Thus, based on the limited data available, we conclude that this burst is not inconsistent with our hypothesis.

Overall, the optical data of GRB 050822, GRB 060729, GRB 060927, and GRB 061222A are not inconsistent with the hypothesis that the prior emission, if it exists, arises from an external shock.  However, due to the lack of optical light curve data for comparison (in the cases of GRB 050822, GRB 060729, and GRB 061222A),  or excess optical flux (in the case of GRB 060927), these bursts do not make a strong case for the hypothesis.  In other words, the results for these GRBs do not constrain or rule out the external shock prior emission hypothesis.

\subsection{Inconsistent Bursts: GRB 050319, GRB 050401, GRB 051109A, and GRB 061121}

The early optical detection of GRB 050319 falls within the thinner part of the green range.  This is consistent with the prediction in which there is one spectral break between the X-ray and optical bands: $\nu_m<\nu_{opt}<\nu_c<\nu_X$.  However, the optical light curve data clearly do not exhibit the smooth, $T_\Delta$-shifted single power law behavior predicted by the external shock prior emission model.  This suggests that, even if the prior emission does originate from an external shock, the optical prior emission component is dominated by other optical components.  Thus, the maximum predicted prior emission optical light curve should lie below the actual optical light curve data.  However, plotting the dashed line below all of the optical data would place the predicted prior emission light curve below the green range, which is not self-consistent.  Thus, the dashed black line is ultimately constrained by the green range in this case.  We must conclude that GRB 050319 is not consistent with the external shock prior emission hypothesis.

The situation is similar for GRB 050401 and GRB 051109A.  Although the early optical detections of these bursts fall within the thinner part of the predicted green range corresponding to $\nu_m<\nu_{opt}<\nu_c<\nu_X$, the shapes of the optical light curves are not consistent with the shape predicted by the external shock prior emission model.  Once again, the predicted prior emission optical light curves are ultimately constrained by the bottom of the green range, and we must conclude that GRB 050401 and GRB 051109A are not consistent with the external shock prior emission hypothesis.

GRB 061121 is the only burst in our sample for which the early optical data fall below the green range.  Consequently, the only way to explain this burst within the external shock prior emission model is to introduce the ad hoc assumption that there are two spectral breaks between the X-ray and optical bands ($\nu_{opt}<\nu_m \lesssim \nu_c < \nu_X$), as indicated by the dash-dotted extension of the green range.  Even after making this ad hoc assumption, there are still issues with GRB 061121.  At first glance the shape of the optical light curve appears promisingly similar to that of the X-ray light curve.  This could be consistent with  Eq. [\ref{X6}] and Figure 2 (given the right set of parameter values), considering that $T_\Delta \sim 2500{~}s$ for this burst.  However, when we actually fit the optical light curve (minus the second data point, which we dismissed as an optical flare) with Eq. [\ref{X1}], the results are not consistent with those of the X-ray light curve fit: $T_\Delta=1100\pm300{~}s$, $\alpha=1.03\pm0.02$, $\tilde{\chi}^{2}~(\nu)=0.83~(35)$. We then tried fitting the optical data, holding the time-shift parameter fixed at the value from the X-ray light curve fit ($T_\Delta=2600{~}s$), but the results still were inconsistent with those of the X-ray light curve fit: $\alpha=1.09\pm0.01$, $\tilde{\chi}^{2}~(\nu)=1.03~(36)$ (this fit is shown as the dashed light blue line in Figure 3). Finally, we tried fitting the optical data with both the time-shift parameter and the decay index fixed at the X-ray values ($T_\Delta=2600{~}s$, $\alpha=1.35$), but this yielded an unacceptable $\tilde{\chi}^{2}~(\nu)=11.2~(37)$. Due to the contrived parameter space required to make this burst consistent with the predictions, as well as the decay index of the optical light curve, we conclude that GRB 061121 is inconsistent with the external shock prior emission hypothesis.

Taken together, the optical data of GRB 050319, GRB 050401, GRB 051109A, and GRB 061121 (which make up half of our sample), severely constrain the external shock prior emission hypothesis.

\section{Conclusion}
In order to test whether or not the observations are consistent with the intriguing hypothesis that the observed X-ray afterglow plateaus are due to an external shock prior emission component, we have compared the prompt (and late, if available) optical emission data with the predicted ranges based on the model. We find that in 4 out of 8 GRBs, the available optical data already impose severe constraints on the hypothesis. In particular, the expected optical flux from the prior external shock model is higher than what the data allow. This suggests that the X-ray emission, if it has a prior emission origin, cannot be synchrotron emission from the external shock. If indeed the shallow-decay X-ray component were from a prior emission component, it would have to be of an internal origin with optical flux suppressed. Future observations before GRB triggers would definitely test the prior emission hypothesis. This might be realized in the SVOM era (see e.g. Paul et al. 2011 for a discussion of the SVOM mission and Zhang 2011 for a discussion of the prospects of detecting prior emission), or perhaps as soon as tomorrow, via serendipitous observations by current observatories.

\bigskip

We acknowledge the anonymous referee whose detailed comments significantly improved this paper.  T.B. thanks Massimiliano de Pasquale, Patricia Schady, and Francisco Virgili for helpful discussion. This work is partially supported by NASA NNX10AD48G and NSF AST-0908362. This research has also made use of data supplied by the UK Swift Science Data Centre at the University of Leicester, and by the NASA/IPAC Extragalactic Database (NED) which is operated by the Jet Propulsion Laboratory, California Institute of Technology, under contract with NASA.

\end{document}